\documentclass[twocolumn,english,pra]{revtex4}
\usepackage[T1]{fontenc}
\usepackage[latin9]{inputenc}
\usepackage{color}
\usepackage{soul}
\usepackage{float}
\usepackage{babel}
\usepackage[normalem]{ulem}
\usepackage{amsmath}
\usepackage{physics}
\usepackage{graphicx}
\usepackage{soul}
\usepackage{tcolorbox}
\usepackage{caption}
\usepackage{subcaption}
\tcbuselibrary{breakable,xparse,skins}
\usepackage{listings}
\usepackage{xcolor}
\definecolor{bg}{RGB}{248,248,248}
\definecolor{keywordcolor}{RGB}{0,128,0}
\definecolor{stringcolor}{RGB}{186,33,33}
\definecolor{commentcolor}{RGB}{64,128,128}
\definecolor{numbercolor}{RGB}{64,160,112}
\definecolor{builtincolor}{RGB}{102,102,255}
\definecolor{framecolor}{RGB}{204,204,204}
\lstnewenvironment{codebox}[1][]
{
    \lstset{#1}
}
{}
\usepackage[unicode=true,pdfusetitle,
 bookmarks=true,bookmarksnumbered=false,bookmarksopen=false,
 breaklinks=false,pdfborder={0 0 1},backref=false,colorlinks=false]
 {hyperref}
 
\makeatletter
\@ifundefined{textcolor}{}
{%
 \definecolor{BLACK}{gray}{0}
 \definecolor{WHITE}{gray}{1}
 \definecolor{RED}{rgb}{1,0,0}
 \definecolor{GREEN}{rgb}{0,1,0}
 \definecolor{BLUE}{rgb}{0,0,1}
 \definecolor{CYAN}{cmyk}{1,0,0,0}
 \definecolor{MAGENTA}{cmyk}{0,1,0,0}
 \definecolor{YELLOW}{cmyk}{0,0,1,0}
}

\usepackage{multirow}
\makeatother
\begin{document}
\title{Accelerating Transpilation in Quantum Machine Learning with Haiqu's Rivet-transpiler}
\author{Aleksander Kaczmarek$^1$, Dikshant Dulal$^{2,3}$}
\affiliation{$^1$ SoftServe Inc, \\$^2$ISAAQ Pte Ltd, $^3$Haiqu}
\email{dikshant@haiqu.ai}
\selectlanguage{english}%
\begin{abstract}
Transpilation is a crucial process in preparing quantum circuits for execution on hardware, transforming virtual gates to match device-specific topology by introducing swap gates and basis gates, and applying optimizations that reduce circuit depth and gate count, particularly for two-qubit gates. As the number of qubits increases, the cost of transpilation escalates significantly, especially when trying to find the optimal layout with minimal noise under the qubit connectivity constraints imposed by device topology. In this work, we use the Rivet transpiler, which accelerates transpilation by reusing previously transpiled circuits. This approach is relevant for cases such as quantum chemistry, where multiple Pauli terms need to be measured by appending a series of rotation gates at the end for non-commuting Paulis, and for more complex cases when quantum circuits need to be modified iteratively, as occurs in quantum layerwise learning. We demonstrate up to $600\%$ improvement in transpilation time for quantum layerwise learning using the Rivet transpiler compared to standard transpilation without reuse.
\end{abstract}
\maketitle
\section{Introduction}
As quantum computing hardware continues to advance, executing complex quantum circuits on these devices poses significant challenges due to hardware-specific constraints such as limited qubit connectivity, gate fidelities, and decoherence times. Addressing these issues requires quantum transpilers that can transform high-level quantum circuits into forms suitable for specific hardware backends \cite{qiskit_sdk, pytket}. The process involves mapping logical qubits to physical qubits, introducing necessary swap gates, and translating gates into the hardware-supported basis sets, all while minimizing circuit depth and reducing gate count. Efficient transpilation is crucial for enhancing fidelity and reducing the execution time of quantum algorithms. Prominent examples of quantum transpilers include IBM's Qiskit \cite{qiskit_sdk}, Quantinuum's tket \cite{pytket}, and BQSKit, each offering unique optimization strategies tailored to different quantum architectures. The effectiveness of these transpilers has been extensively studied, highlighting their impact on overall circuit performance and making the choice of transpiler a critical factor in practical quantum applications \cite{comparativestudyquantumtranspilers}.

\subsection{Transpilation}
A critical challenge in deploying quantum algorithms on physical devices is the translation of high-level quantum circuits into hardware-compatible instructions, a process known as \emph{transpilation}. Transpilation must account for device-specific constraints, including qubit connectivity, native gate sets, and operational errors, to optimize circuit performance and reliability. This process involves several interconnected stages. For introduction, we will discuss how Qiskit performs transpilation \cite{qiskit_sdk}. Initially, the circuit undergoes \textbf{initialization}, where gate representations are standardized, and custom instructions are unrolled into basic quantum operations. Following this, a \textbf{layout} stage maps the circuit's virtual qubits to the physical qubits of the hardware, ensuring efficient utilization of resources. To address qubit connectivity limitations inherent in the hardware, the \textbf{routing} stage introduces SWAP gates, rearranging qubits to satisfy these constraints. The circuit is then subjected to \textbf{translation}, converting its gates into the device's native gate set for compatibility. An \textbf{optimization} phase ensues, where circuit depth and gate count are reduced through iterative refinement techniques, enhancing performance and mitigating error accumulation. Finally, \textbf{scheduling} aligns gate operations temporally according to hardware-specific timing requirements, ensuring coherent execution of the quantum circuit.
\begin{figure}[h]
    \centering
    \includegraphics[width= \linewidth]{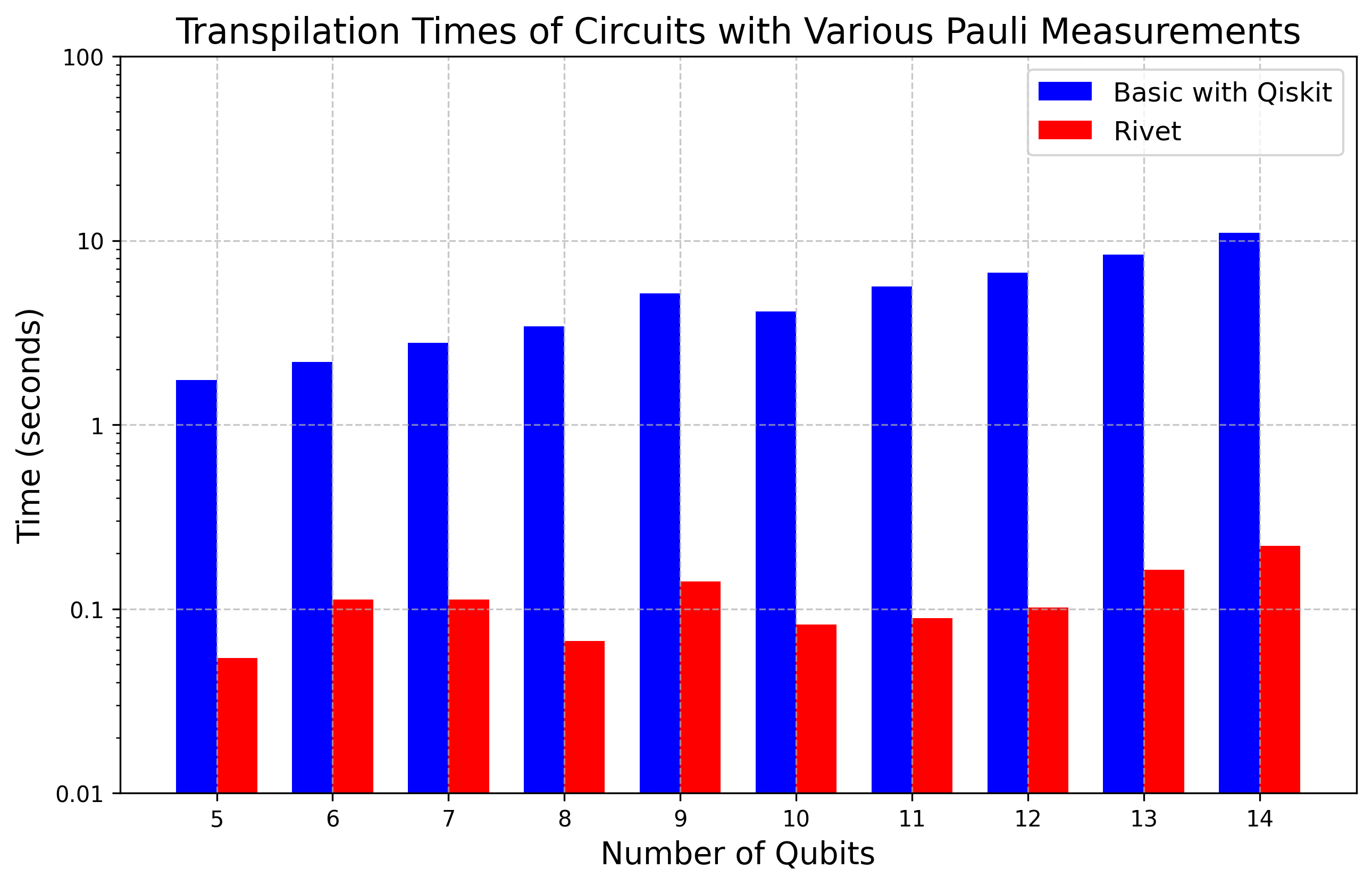}
    \caption{Warm up example comparing transpilation time for a random input circuit measured in 10 randomly generated Pauli bases.}
    \label{fig:warm_up_eg}
\end{figure}
\subsection{Rivet Transpiler}
Conventional transpilation methods operate on individual circuits without taking into account the broader experimental structure. This often leads to redundant transpilation steps when multiple, structurally similar circuits are required, as in iterative algorithms or experiments involving repeated measurements with varying bases. Haiqu's Rivet transpiler addresses this inefficiency by caching and reusing transpiled subcircuits, which is particularly beneficial in scenarios such as state tomography, where reconstructing an $n$-qubit density matrix requires $3^n$ distinct measurement circuits, and shadow tomography\cite{classicalshadows}, where the same circuit is measured in multiple bases to evaluate many-body properties of interest. Furthermore, Rivet becomes useful in the context of variational quantum algorithms, particularly in quantum chemistry, where the optimization loop involves repeatedly measuring the same circuit in multiple Pauli bases. Instead of transpiling each circuit independently, Rivet transpiles the common state preparation circuit once and reuses it across all configurations by appending the basis rotations as needed. This reuse capability drastically reduces transpilation time, making Rivet well-suited for experiments with high transpilation demands.

The advantages of Rivet become even more apparent in the context of quantum machine learning, where the transpilation overhead is further amplified. Quantum layerwise learning (LL) is an example of a machine learning strategy that incrementally adds new parameterized quantum layers to the circuit, requiring each version of the circuit to be transpiled before deployment. Using conventional transpilers, this iterative modification can lead to substantial delays, particularly as the number of layers or qubits increases. Rivet mitigates this issue by caching previously transpiled versions of the circuit and applying only the incremental changes, resulting in speedups of over 100x for larger quantum systems compared to standard transpilers.

The remainder of this paper is organized as follows: Section \ref{section:Implementation} introduces the core architecture of the Rivet transpiler and discusses its integration with the Layerwise Learning framework. In Section \ref{section:Experiments}, we present experimental results demonstrating the effectiveness of Rivet in reducing transpilation time for binary classification tasks using quantum neural networks. Finally, Section \ref{section:Conclusions} summarizes our findings and provides insights into optimizing transpilation strategies for iterative quantum algorithms.
\section{Implementation}
\label{section:Implementation}
\subsection{Warm Up Example}
\label{subsection:warmup}
To consider a simple example of how to use rivet, we take  \verb|EfficientSU2|
ansatz from qiskit \cite{qiskit2024} which is a commonly used hardware efficient ansatz for training VQE problems. This is to reflect the example of post-rotations done for Variational Quantum eigensolvers,  State Tomography where the same circuit is measured different bases. We then generate 10 random paulis each of which is to be measured on the input circuit by applying post-rotations. 

First we import the mentioned \verb|transpile_right| from rivet with the following command.

\begin{codebox}[language=Python]
from rivet_transpiler import transpile_right
\end{codebox}

Then we import other necessary functions from qiskit for transpilation and generating circuit, and running on a fake device in qiskit.

\begin{codebox}[language=Python]
from qiskit.circuit.random import random_circuit
from qiskit_ibm_runtime.fake_provider import FakeKolkataV2
from qiskit import QuantumCircuit, transpile
from rivet_transpiler import transpile_right
\end{codebox}

We then take \verb|FakeKolkataV2| which is a backend that has topology and basis gates along with noise models that mimicks the noise in real device. We choose a 6 qubit random circuit from qiskit using the function \verb|random_circuit|. 

\begin{codebox}[language=Python]
device = FakeKolkataV2()
NQUBITS = 6
INPUT_QC = random_circuit(NQUBITS, depth=10)
PAULI_STR = "IXYIYX"
\end{codebox}

We then prepare a rotation circuit for measuring in pauli bases. If the qubit is to be measured in computational basis then it is left unchanged, for X we apply the Hadamard (H) gate and for Y we apply Hadamard and then $S^\dagger$ gate. The helper functions for preparing the rotation circuit and generating random paulis is provided in appendix.

\begin{codebox}[language=Python]
def create_rotation_circuit(
    circ: QuantumCircuit, pauli_str: str
) -> QuantumCircuit:
    qc = QuantumCircuit(circ.num_qubits)
    assert len(pauli_str) == circ.num_qubits, "Pauli string does not match number of qubits."
    for i, pauli in enumerate(pauli_str[::-1]):
        if pauli == "X":
            qc.h(i)
        elif pauli == "Y":
            qc.sdg(i)
            qc.h(i)
    return qc
\end{codebox}

Now we will transpile each of the input circuit with rotation one at a time in qiskit with the above for loop.

To show how basic transpilation done, we first combine the circuits and then measure the combined circuit as follows:

\begin{codebox}[language=Python]
rotation_qc = create_rotation_circuit(INPUT_QC, PAULI_STR)
basic_qc = QuantumCircuit(NQUBITS)
basic_qc.append(INPUT_QC, range(NQUBITS))
basic_qc.append(rotation_qc, range(NQUBITS))
basic_qc.measure_all()
basic_transpiled_qc = transpile(basic_qc, device, optimization_level = 3, seed_transpiler=42)
\end{codebox}

To transpile with rivet we first remove the measurements of the transpiled circuit \verb|TRANSPILED_QC| and measure the right circuit which will be resued during with rivet.

Now we simply use the \verb|transpiled_right| function as follows:
\begin{codebox}[language=Python]
rot_qc_w_meas = rotation_qc.measure_all(inplace=False)
t_qc = TRANSPILED_QC.remove_final_measurements(
inplace=False)
rivet_transpiled_circuit = transpile_right(t_qc, rot_qc_w_meas, device)
\end{codebox}

Note that we first need to measure the untranspiled right circuit that needs to be combined with the transpiled left circuit. We can check that the outcomes of the circuits are the same by checking the fidelity.
\begin{codebox}[language=Python]
from qiskit_aer import AerSimulator
from qiskit.quantum_info import hellinger_fidelity
ideal_dev = AerSimulator()
basic_qc_res = (
    ideal_dev.run(basic_transpiled_qc, shots=100_000).result().get_counts()
)
rivet_qc_res = (
    ideal_dev.run(rivet_transpiled_circuit, shots=100_000).result().get_counts()
)
print(hellinger_fidelity(basic_qc_res, rivet_qc_res))
\end{codebox}

\subsection{Layerwise Learning}
Layerwise Learning (LL) was  proposed \cite{layerwise_learning_paper} as a method for training quantum machine learning models to address the problem of barren plateaus \cite{barrenplateaus}, which is analogous to the vanishing gradient problem in classical neural networks. In parameterized quantum circuits (PQCs), the barren plateau phenomenon arises when the gradients of the cost function with respect to the circuit parameters approach zero exponentially as the number of qubits or the circuit depth increases \cite{barrenplateaus}. This occurs because the optimization landscape of the PQC becomes overwhelmingly flat, with most regions of the parameter space exhibiting near-zero gradients. As a result, randomly initialized PQCs are highly likely to start in these flat regions, making it computationally infeasible to find optimal parameters through gradient-based methods. 

Mathematically, the presence of barren plateaus is characterized by the variance of the gradient scaling exponentially with the number of qubits, i.e., $\text{Var}(\partial C / \partial \theta) \propto 2^{-n}$, where $C$ is the cost function, $\theta$ is a parameter of the circuit, and $n$ is the number of qubits. As $n$ increases, the gradients become exponentially smaller, leading to a landscape where the optimization procedure is unable to make meaningful updates to the parameters. The problem is exacerbated as circuit depth increases, as deeper circuits tend to explore a larger portion of the Hilbert space, resulting in a uniform distribution over the parameter space that induces flat cost surfaces. In Noisy Intermediate-Scale Quantum (NISQ) devices, noise further contributes to the emergence of barren plateaus. Quantum noise, such as decoherence and gate errors, introduces randomness into the circuit evolution, effectively smearing out the gradients and creating noise-induced barren plateaus\cite{noiseinducedBI}. Consequently, overcoming barren plateaus is essential for scalable quantum machine learning and variational quantum algorithms on NISQ devices.

To overcome these limitations, LL was designed to simultaneously build and optimize PQCs in a stepwise manner, thereby avoiding barren plateaus by maintaining larger gradient magnitudes during training. The training process is divided into two phases. In Phase 1, the circuit is constructed iteratively by adding layers in multiple steps. At each step, a predetermined number of layers is appended, and the circuit is optimized to find suitable parameters. Once the circuit reaches the desired depth and initial parameters are established, Phase 2 begins. During Phase 2, the PQC is divided into partitions, and each partition is optimized sequentially while keeping the remaining parameters fixed. This partitioned training strategy helps prevent barren plateaus by ensuring low-depth circuits are trained in each step, maintaining a higher gradient magnitude.

In this work, we emphasize the efficiency improvements achieved in Phase 1 using the Rivet transpiler. Typically, each layer addition step would require a full circuit transpilation on hardware, leading to significant overhead, especially for deep circuits. However, Rivet's \texttt{transpile\_right} function allows for incremental transpilation, reusing previously transpiled segments and only transpiling the newly added layers and stiching the transpiled circuit together. This significantly reduces transpilation time, making LL more practical for training deep PQCs. Additionally, ensuring that the data encoding circuits are properly aligned with the circuit topology is crucial for seamless integration during layer-by-layer growth, which is efficiently handled by Rivet's transpilation capabilities.


\subsection{Transpilation with rivet for Layerwise Learning}

Efficient transpilation is crucial in quantum machine learning, particularly when training parameterized quantum circuits (PQCs) using advanced methods like Layerwise Learning (LL). The Rivet transpiler offers the \verb|transpile_right| function, which can append trainable PQCs to pre-transpiled encoded data samples. This approach is especially beneficial when the data input circuit is deep and challenging to transpile - such as circuits utilizing \verb|prepare_state|\cite{qiskit2024} or the ZZFeatureMap\cite{qiskit2024} - due to the limited connectivity imposed by the hardware's coupling map.

In the LL training approach, the PQC is simultaneously built and trained during Phase 1, requiring transpiled versions of the circuit at each consecutive layer addition step. The use of \verb|transpile_right| with Rivet becomes advantageous here, as it allows for the reuse of previously transpiled circuit components, significantly reducing transpilation time across iterations.

We conducted experiments comparing naive (basic) transpilation and Rivet transpilation times for circuits required to train a quantum machine learning model using the LL approach with three distinct data encoding strategies:

\begin{enumerate}
    \item \textbf{Angle Encoding}
    \item \textbf{Amplitude Encoding}
    \item \textbf{ZZFeatureMap Encoding} (discussed in Section~\ref{sec:ZZFeatureMap})
\end{enumerate}

Our analysis focused on the scaling of transpilation time concerning various circuit sizes and numbers of layers, both of which correspond to different counts of trainable parameters in the PQC.

\subsubsection{Angle Encoding}
\label{sec:angle_encoding}
Angle Encoding maps classical data into quantum states using parameterized single-qubit rotation gates. In our experiments, we trained PQCs of varying depths using the LL approach, incrementally adding layers to the circuit.

\begin{figure}[H]
\centering\includegraphics[scale=0.4]{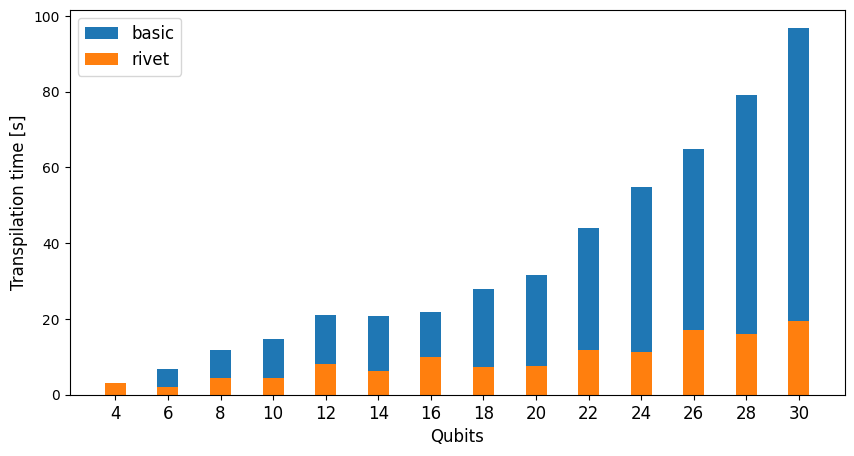}
\caption{Comparison of transpilation times between basic and Rivet methods for training a 20-layer PQC using Angle Encoding with Layerwise Learning, adding 2 layers at each of 10 steps.}
\label{fig:transpilation_time_las10}
\end{figure}

Figure~\ref{fig:transpilation_time_las10} presents the transpilation times for training a PQC with 20 layers. The LL approach added 2 layers at each step over 10 steps. Without Rivet, transpilation took approximately five times longer than with Rivet. This substantial reduction demonstrates Rivet's efficiency in reusing transpiled components.

\begin{figure}[H]
\centering
\includegraphics[scale=0.4]{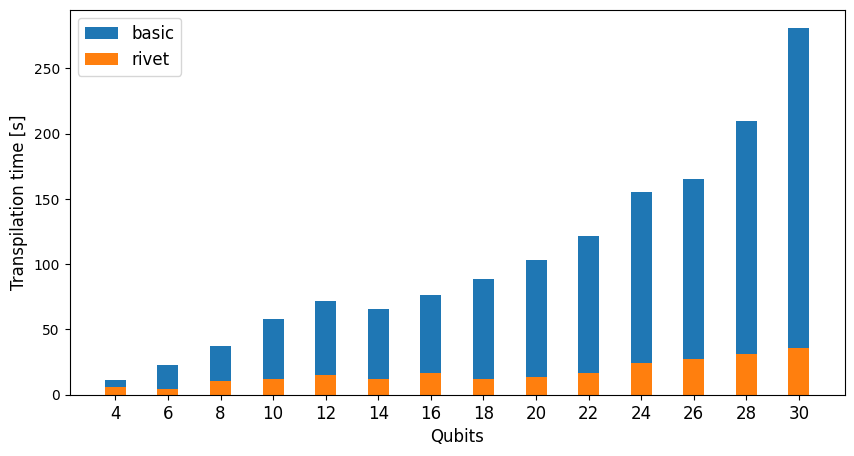}
\caption{Comparison of transpilation times between basic and Rivet methods for training a 40-layer PQC using Angle Encoding with Layerwise Learning, adding 2 layers at each of 20 steps.}
\label{fig:transpilation_time_las20}
\end{figure}

For deeper circuits, Rivet's benefits become more pronounced. Figure~\ref{fig:transpilation_time_las20} shows transpilation times for a 40-layer PQC, with layers added in 20 steps of 2 layers each. Here, Rivet reduced transpilation time by a factor of eight compared to the basic method. Both experiments used simple data input layers composed of single-qubit rotation gates.

These results indicate that Rivet's \verb|transpile_right| function significantly enhances transpilation efficiency, particularly as circuit depth increases. By avoiding redundant computations through component reuse, Rivet streamlines the training workflow.

\subsubsection{Amplitude Data Encoding}
\label{subsubsec:statevector_data_encoding}
Amplitude Encoding allows for encoding $2^n$ classical features into the amplitudes of an $n$-qubit quantum state. In Qiskit \cite{qiskit2024}, this is implemented via the \verb|prepare_state| \cite{qiskit2024} method. Unlike Angle Encoding or ZZFeatureMap Encoding, Amplitude Encoding does not yield a parameterized circuit, and the transpilation time can vary based on the feature values due to the dynamic structure of the generated state preparation circuit.

\begin{figure}[h]
\centering
\includegraphics[scale=0.4]{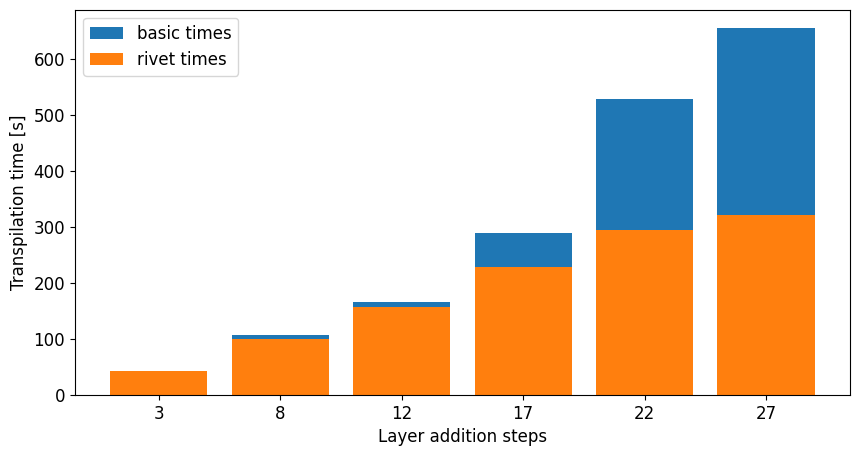}
\caption{Transpilation times for training a 4-qubit PQC with Amplitude Encoding using Layerwise Learning. Four layers are added at each step, comparing basic and Rivet methods.}
\label{fig:transpilation_time_ps_4qubits}
\end{figure}

\begin{figure}[h]
\centering
\includegraphics[scale=0.4]{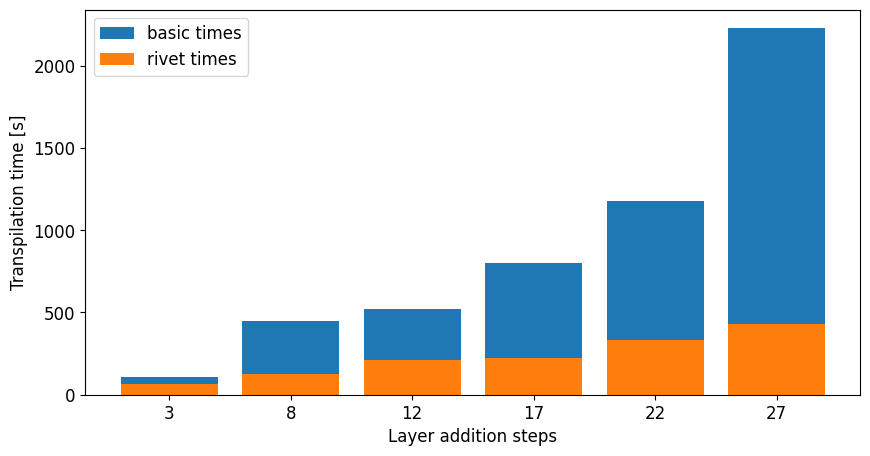}
\caption{Transpilation times for training a 6-qubit PQC with Amplitude Encoding using Layerwise Learning. Four layers are added at each step, comparing basic and Rivet methods.}
\label{fig:transpilation_time_ps_6qubits}
\end{figure}

We experimented with PQCs of 4 and 6 qubits using Amplitude Encoding. The MNIST digits dataset was rescaled from $28 \times 28$ pixels to $4 \times 4$ pixels for the 4-qubit case and to $8 \times 8$ pixels for the 6-qubit case, encoding 16 and 64 features, respectively. Figure~\ref{fig:transpilation_time_ps_4qubits} displays transpilation times for a 4-qubit PQC trained using the LL approach, adding 4 layers at each step. Rivet reduced the transpilation time compared to the basic method, illustrating its effectiveness even with variable data encoding circuits.
In Figure~\ref{fig:transpilation_time_ps_6qubits}, we present results for a 6-qubit PQC with different number of layers. In particular, for a 6 qubit PQC with 648 parameters, rivet achieved up to a fourfold reduction in transpilation time, highlighting its scalability and efficiency for larger circuits with more parameters.
\subsubsection{Discussion}
Our findings demonstrate that Rivet's \verb|transpile_right| function offers substantial improvements in transpilation time across different data encoding strategies and PQC configurations. The efficiency gains are more significant for deeper circuits and larger qubit systems, which are common in practical quantum machine learning applications.
The ability to reuse transpiled components not only accelerates the transpilation process but also contributes to resource optimization, which is crucial when working with limited computational capabilities or when rapid iteration is necessary. This is particularly relevant in the LL training approach, where circuits are incrementally expanded and retrained, making traditional transpilation methods computationally expensive.
Overall, Rivet's approach aligns with the needs of scalable quantum computing, providing a practical solution to the challenges of transpiling increasingly complex quantum circuits. Its integration into the quantum software stack can enhance the development and deployment of quantum algorithms, especially in machine learning tasks where iterative circuit modifications are frequent.
\section{Experiments}
\label{section:Experiments}

\subsubsection{Angle Encoding}

\begin{figure}[h]
\hspace{-0.5cm}
\centering
\includegraphics[scale=0.35]{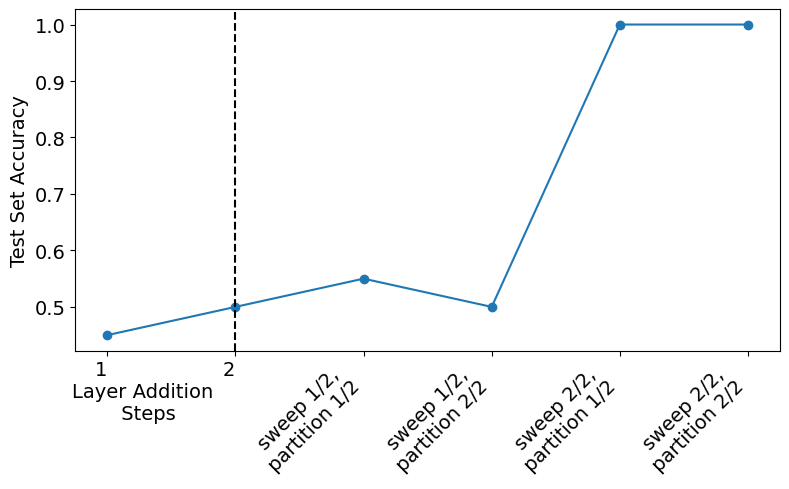}
\caption{Test accuracy during training phases. In Phase One, accuracy is calculated after each added layer completes training. In Phase Two, it is calculated after training each partition.}
\label{fig:testacc_iris}
\end{figure}

We conducted a binary classification learning experiment using the Iris dataset from scikit-learn, following the example in \cite{iris_qiskit}. To create a binary classification problem, we performed data preprocessing by removing one of the classes from the dataset. Using this data, we trained a binary classification model based on a Parameterized Quantum Circuit (PQC) with measurement on the last qubit, employing the Layerwise Learning (LL) approach as described in \cite{layerwise_learning_paper}. Unlike \cite{layerwise_learning_paper}, which encodes features using X gates, we utilized angle encoding (see Section~\ref{sec:angle_encoding}) with parameterized RX gates and our custom quantum neural network (QNN) class (i.e., not using Qiskit Machine Learning as in \cite{iris_qiskit}). As shown in Figure~\ref{fig:testacc_iris}, a PQC with 4 qubits and 6 layers (totaling 24 trainable parameters) achieved 100\% test accuracy using the LL approach.

\begin{figure}[h]
\hspace{-0.5cm}
\centering
\includegraphics[scale=0.35]{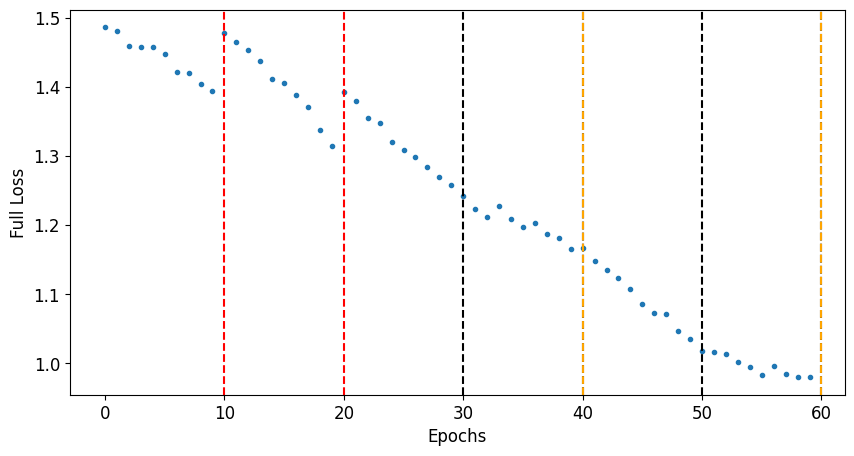}
\caption{Loss function values during training phases. Layer additions are marked with red dashed lines; each line indicates a layer addition step of adding 3 layers. In Phase Two, we perform 2 sweeps over 2 equal partitions. The orange dashed line marks the end of training of the first partition, and the black dashed line marks the end of training over the second partition.}
\label{fig:loss_iris}
\end{figure}

\subsubsection{Amplitude Data Encoding}
\begin{figure}[h]
\centering
\includegraphics[scale=0.35]{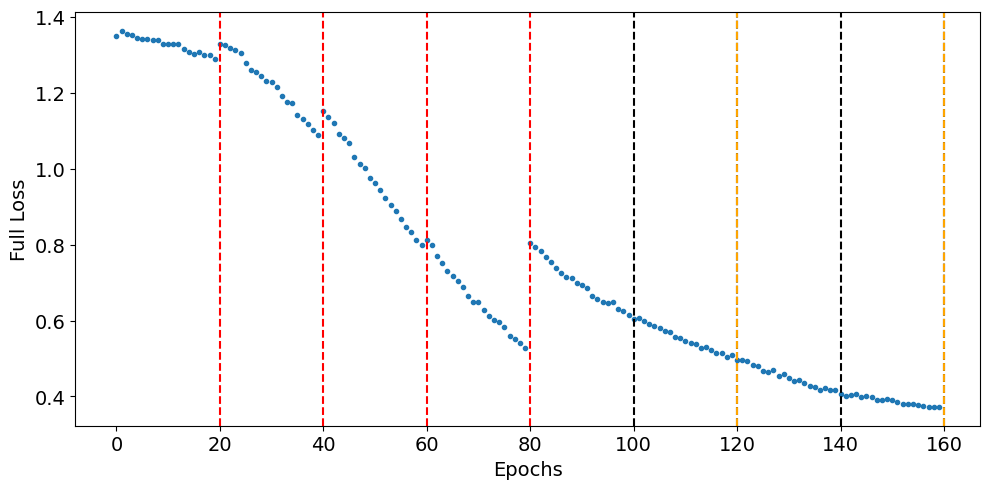}
\caption{Combined loss function from Phase One and Phase Two. Red lines separate each layer addition step. Black and orange lines separate partition training (black for Partition 1 and orange for Partition 2) in Phase Two.}
\label{fig:loss}
\end{figure}

\begin{figure}[h]
\centering
\includegraphics[scale=0.35]{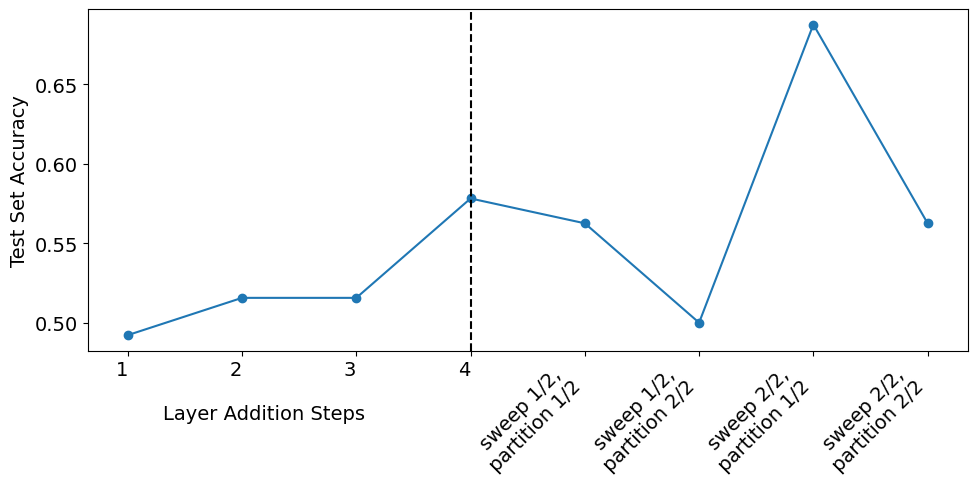}
\caption{Test accuracy during training phases. In Phase One, accuracy is calculated after each added layer completes training. In Phase Two, it is calculated after training each partition.}
\label{fig:testacc}
\end{figure}

\begin{figure}[h]
\centering
\includegraphics[scale=0.4]{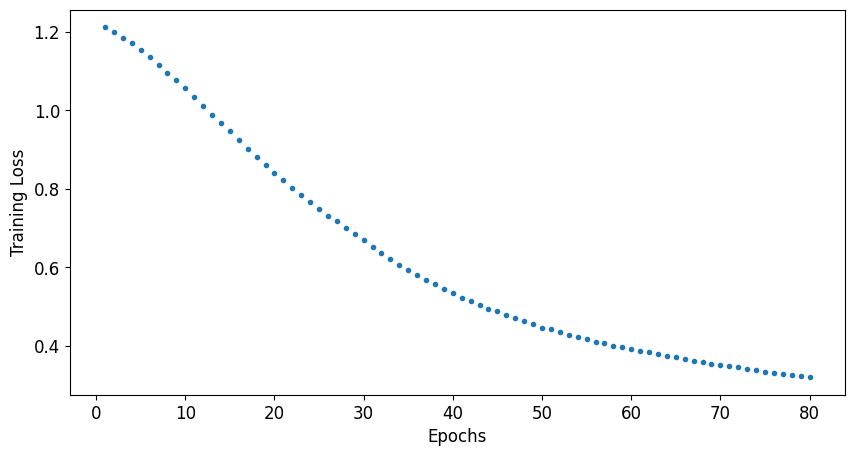}
\caption{Loss function evaluated at each epoch during regular training of a 12-layer PQC.}
\label{fig:loss_regular}
\end{figure}

\begin{figure}[h]
\centering
\includegraphics[scale=0.4]{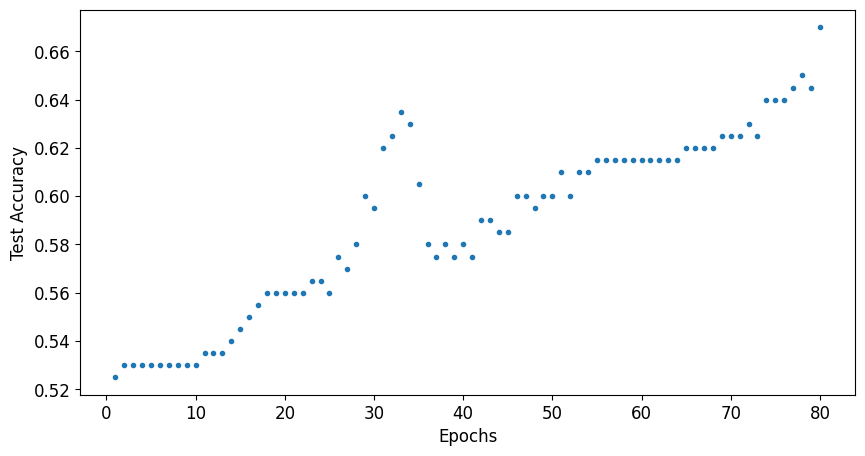}
\caption{Test accuracy evaluated at each epoch during regular training of a 12-layer PQC.}
\label{fig:test_acc_regular}
\end{figure}

We also performed a binary classification learning experiment with PQCs trained using the Layerwise Learning approach. The classification task is the same as presented in \cite{layerwise_learning_paper}, aiming to train the PQC to distinguish between handwritten digits 3 and 6. Following the GitHub repository \cite{layerwise_learning_paper} associated with the paper, the images were downsized to a $4 \times 4$ matrix, resulting in 16 features per sample. Since we are using Amplitude Encoding, we employed the \verb|prepare_state| method for the data input layer instead of the layer of X gates used in \cite{layerwise_learning_paper}. Thus, we used a PQC with only 4 qubits to encode the same number of features (16) as in \cite{layerwise_learning_paper}. We compared a 12-layer PQC trained with the LL approach to the regular training approach (where all layers are trained at once on a fully built PQC). For the LL training, we chose 4 layer addition steps of 3 layers each and performed 2 sweeps over 2 partitions in Phase Two. The test loss and test accuracy of the LL approach are presented in Figures~\ref{fig:loss} and \ref{fig:testacc}.
We found that for this task, the PQC with the same number of layers achieved comparable results to the regular training approach, as presented in Figures~\ref{fig:loss_regular} and \ref{fig:test_acc_regular}. These results are also on par with those presented in \cite{layerwise_learning_paper}.
\section{Conclusion}
\label{section:Conclusions}
The Rivet transpiler is a novel solution designed to address the inefficiencies in quantum circuit compilation, particularly in quantum machine learning and other iterative processes. Quantum machine learning, by its nature, is a transpilation-heavy process as each data sample must be encoded into a quantum circuit, which introduces a significant computational burden. This burden can be mitigated through the use of appropriate data encoding strategies and transpilation optimization techniques. The Rivet transpiler specifically targets scenarios where circuits share common structures, allowing for substantial reductions in transpilation time by reusing previously compiled segments and minimizing the need for recompilation.

Different data encoding strategies vary in their impact on transpilation efficiency, and Rivet's flexibility allows it to adapt to these variations. Parameterized quantum data encoding schemes, such as Angle encoding (which employs single-qubit rotation layers) or the ZZFeatureMap (discussed in Section~\ref{sec:ZZFeatureMap}), encode dataset features as circuit parameters without specifying their values during the transpilation step. This approach enables the transpilation of a single parameterized circuit, which can then be executed with different feature values at runtime. While efficient, these methods come with a trade-off in terms of expressibility, as each feature requires a separate qubit, leading to potentially large circuits when the number of features increases.

Conversely, more advanced data encoding techniques, such as \verb|prepare_state| in Qiskit, are capable of encoding $2^n$ features into the statevector of $n$ qubits, significantly improving the efficiency of feature utilization in the Hilbert space. However, the cost is that these circuits need to be transpiled independently for each dataset sample, resulting in deeper circuits and longer transpilation times for larger datasets. Rivet's reuse mechanism is particularly beneficial in such cases, as it enables the pre-transpilation of shared components, making the preparation and execution of these circuits faster and more scalable.

The advantages of Rivet are most apparent in the Layerwise Learning (LL) approach, where the circuit is constructed and optimized incrementally, layer by layer. In Phase One of LL, where new layers are sequentially added to the Parameterized Quantum Circuit (PQC), Rivet significantly reduces the transpilation overhead by reusing previously transpiled segments and only transpiling the newly appended layers. This incremental transpilation not only saves time but also ensures that the overall training process remains efficient, even for deep circuits with a large number of layers.

Our experiments demonstrate that the use of the Rivet transpiler leads to substantial reductions in transpilation time across various data encoding strategies and PQC configurations. This efficiency is crucial for scaling quantum machine learning algorithms to handle larger datasets and more complex models. The integration of Rivet into the quantum software stack facilitates quicker experimentation and development cycles, contributing to the broader goal of practical and efficient quantum computing. By reducing the compilation overhead, Rivet empowers researchers to focus on algorithm development and problem-solving, accelerating progress in quantum machine learning and other computationally intensive quantum applications.

\textbf{Code and data availability:} The program used to reproduce training results with layerwise learning and compare transpilation times is available upon request. This program was developed using the open-source package \href{https://github.com/haiqu-ai/rivet}{rivet}.
\bibliographystyle{apsrev4-1}
\bibliography{citations}

\begin{thebibliography}{9}%
\makeatletter
\providecommand \@ifxundefined [1]{%
 \@ifx{#1\undefined}
}%
\providecommand \@ifnum [1]{%
 \ifnum #1\expandafter \@firstoftwo
 \else \expandafter \@secondoftwo
 \fi
}%
\providecommand \@ifx [1]{%
 \ifx #1\expandafter \@firstoftwo
 \else \expandafter \@secondoftwo
 \fi
}%
\providecommand \natexlab [1]{#1}%
\providecommand \enquote  [1]{``#1''}%
\providecommand \bibnamefont  [1]{#1}%
\providecommand \bibfnamefont [1]{#1}%
\providecommand \citenamefont [1]{#1}%
\providecommand \href@noop [0]{\@secondoftwo}%
\providecommand \href [0]{\begingroup \@sanitize@url \@href}%
\providecommand \@href[1]{\@@startlink{#1}\@@href}%
\providecommand \@@href[1]{\endgroup#1\@@endlink}%
\providecommand \@sanitize@url [0]{\catcode `\\12\catcode `\$12\catcode `\&12\catcode `\#12\catcode `\^12\catcode `\_12\catcode `\%12\relax}%
\providecommand \@@startlink[1]{}%
\providecommand \@@endlink[0]{}%
\providecommand \url  [0]{\begingroup\@sanitize@url \@url }%
\providecommand \@url [1]{\endgroup\@href {#1}{\urlprefix }}%
\providecommand \urlprefix  [0]{URL }%
\providecommand \Eprint [0]{\href }%
\providecommand \doibase [0]{http://dx.doi.org/}%
\providecommand \selectlanguage [0]{\@gobble}%
\providecommand \bibinfo  [0]{\@secondoftwo}%
\providecommand \bibfield  [0]{\@secondoftwo}%
\providecommand \translation [1]{[#1]}%
\providecommand \BibitemOpen [0]{}%
\providecommand \bibitemStop [0]{}%
\providecommand \bibitemNoStop [0]{.\EOS\space}%
\providecommand \EOS [0]{\spacefactor3000\relax}%
\providecommand \BibitemShut  [1]{\csname bibitem#1\endcsname}%
\let\auto@bib@innerbib\@empty
\bibitem [{\citenamefont {Javadi-Abhari}\ \emph {et~al.}(2024{\natexlab{a}})\citenamefont {Javadi-Abhari}, \citenamefont {Treinish}, \citenamefont {Krsulich}, \citenamefont {Wood}, \citenamefont {Lishman}, \citenamefont {Gacon}, \citenamefont {Martiel}, \citenamefont {Nation}, \citenamefont {Bishop}, \citenamefont {Cross} \emph {et~al.}}]{qiskit_sdk}%
  \BibitemOpen
  \bibfield  {author} {\bibinfo {author} {\bibfnamefont {A.}~\bibnamefont {Javadi-Abhari}}, \bibinfo {author} {\bibfnamefont {M.}~\bibnamefont {Treinish}}, \bibinfo {author} {\bibfnamefont {K.}~\bibnamefont {Krsulich}}, \bibinfo {author} {\bibfnamefont {C.~J.}\ \bibnamefont {Wood}}, \bibinfo {author} {\bibfnamefont {J.}~\bibnamefont {Lishman}}, \bibinfo {author} {\bibfnamefont {J.}~\bibnamefont {Gacon}}, \bibinfo {author} {\bibfnamefont {S.}~\bibnamefont {Martiel}}, \bibinfo {author} {\bibfnamefont {P.~D.}\ \bibnamefont {Nation}}, \bibinfo {author} {\bibfnamefont {L.~S.}\ \bibnamefont {Bishop}}, \bibinfo {author} {\bibfnamefont {A.~W.}\ \bibnamefont {Cross}},  \emph {et~al.},\ }\href@noop {} {\enquote {\bibinfo {title} {Quantum computing with qiskit},}\ } (\bibinfo {year} {2024}{\natexlab{a}})\BibitemShut {NoStop}%
\bibitem [{\citenamefont {Sivarajah}\ \emph {et~al.}(2020)\citenamefont {Sivarajah}, \citenamefont {Dilkes}, \citenamefont {Cowtan}, \citenamefont {Simmons}, \citenamefont {Edgington},\ and\ \citenamefont {Duncan}}]{pytket}%
  \BibitemOpen
  \bibfield  {author} {\bibinfo {author} {\bibfnamefont {S.}~\bibnamefont {Sivarajah}}, \bibinfo {author} {\bibfnamefont {S.}~\bibnamefont {Dilkes}}, \bibinfo {author} {\bibfnamefont {A.}~\bibnamefont {Cowtan}}, \bibinfo {author} {\bibfnamefont {W.}~\bibnamefont {Simmons}}, \bibinfo {author} {\bibfnamefont {A.}~\bibnamefont {Edgington}}, \ and\ \bibinfo {author} {\bibfnamefont {R.}~\bibnamefont {Duncan}},\ }\href@noop {} {\enquote {\bibinfo {title} {t| ket>: a retargetable compiler for nisq devices},}\ } (\bibinfo {year} {2020})\BibitemShut {NoStop}%
\bibitem [{\citenamefont {Louamri}\ \emph {et~al.}(2024)\citenamefont {Louamri}, \citenamefont {eddine Belaloui}, \citenamefont {Tounsi},\ and\ \citenamefont {Rouabah}}]{comparativestudyquantumtranspilers}%
  \BibitemOpen
  \bibfield  {author} {\bibinfo {author} {\bibfnamefont {M.~M.}\ \bibnamefont {Louamri}}, \bibinfo {author} {\bibfnamefont {N.}~\bibnamefont {eddine Belaloui}}, \bibinfo {author} {\bibfnamefont {A.}~\bibnamefont {Tounsi}}, \ and\ \bibinfo {author} {\bibfnamefont {M.~T.}\ \bibnamefont {Rouabah}},\ }\href {https://arxiv.org/abs/2406.06836} {\enquote {\bibinfo {title} {Comparative study of quantum transpilers: Evaluating the performance of qiskit-braket-provider, qbraid-sdk, and pytket extensions},}\ } (\bibinfo {year} {2024}),\ \Eprint {http://arxiv.org/abs/2406.06836} {arXiv:2406.06836 [quant-ph]} \BibitemShut {NoStop}%
\bibitem [{\citenamefont {Huang}\ \emph {et~al.}(2022)\citenamefont {Huang}, \citenamefont {Kueng}, \citenamefont {Torlai}, \citenamefont {Albert},\ and\ \citenamefont {Preskill}}]{classicalshadows}%
  \BibitemOpen
  \bibfield  {author} {\bibinfo {author} {\bibfnamefont {H.-Y.}\ \bibnamefont {Huang}}, \bibinfo {author} {\bibfnamefont {R.}~\bibnamefont {Kueng}}, \bibinfo {author} {\bibfnamefont {G.}~\bibnamefont {Torlai}}, \bibinfo {author} {\bibfnamefont {V.~V.}\ \bibnamefont {Albert}}, \ and\ \bibinfo {author} {\bibfnamefont {J.}~\bibnamefont {Preskill}},\ }\href@noop {} {\bibfield  {journal} {\bibinfo  {journal} {Science}\ }\textbf {\bibinfo {volume} {377}},\ \bibinfo {pages} {eabk3333} (\bibinfo {year} {2022})}\BibitemShut {NoStop}%
\bibitem [{\citenamefont {Javadi-Abhari}\ \emph {et~al.}(2024{\natexlab{b}})\citenamefont {Javadi-Abhari}, \citenamefont {Treinish}, \citenamefont {Krsulich}, \citenamefont {Wood}, \citenamefont {Lishman}, \citenamefont {Gacon}, \citenamefont {Martiel}, \citenamefont {Nation}, \citenamefont {Bishop}, \citenamefont {Cross}, \citenamefont {Johnson},\ and\ \citenamefont {Gambetta}}]{qiskit2024}%
  \BibitemOpen
  \bibfield  {author} {\bibinfo {author} {\bibfnamefont {A.}~\bibnamefont {Javadi-Abhari}}, \bibinfo {author} {\bibfnamefont {M.}~\bibnamefont {Treinish}}, \bibinfo {author} {\bibfnamefont {K.}~\bibnamefont {Krsulich}}, \bibinfo {author} {\bibfnamefont {C.~J.}\ \bibnamefont {Wood}}, \bibinfo {author} {\bibfnamefont {J.}~\bibnamefont {Lishman}}, \bibinfo {author} {\bibfnamefont {J.}~\bibnamefont {Gacon}}, \bibinfo {author} {\bibfnamefont {S.}~\bibnamefont {Martiel}}, \bibinfo {author} {\bibfnamefont {P.~D.}\ \bibnamefont {Nation}}, \bibinfo {author} {\bibfnamefont {L.~S.}\ \bibnamefont {Bishop}}, \bibinfo {author} {\bibfnamefont {A.~W.}\ \bibnamefont {Cross}}, \bibinfo {author} {\bibfnamefont {B.~R.}\ \bibnamefont {Johnson}}, \ and\ \bibinfo {author} {\bibfnamefont {J.~M.}\ \bibnamefont {Gambetta}},\ }\href {\doibase 10.48550/arXiv.2405.08810} {\enquote {\bibinfo {title} {Quantum computing with {Q}iskit},}\ } (\bibinfo {year} {2024}{\natexlab{b}}),\ \Eprint {http://arxiv.org/abs/2405.08810} {arXiv:2405.08810
  [quant-ph]} \BibitemShut {NoStop}%
\bibitem [{\citenamefont {Skolik}\ \emph {et~al.}(2021)\citenamefont {Skolik}, \citenamefont {McClean}, \citenamefont {Mohseni}, \citenamefont {van~der Smagt},\ and\ \citenamefont {Leib}}]{layerwise_learning_paper}%
  \BibitemOpen
  \bibfield  {author} {\bibinfo {author} {\bibfnamefont {A.}~\bibnamefont {Skolik}}, \bibinfo {author} {\bibfnamefont {J.~R.}\ \bibnamefont {McClean}}, \bibinfo {author} {\bibfnamefont {M.}~\bibnamefont {Mohseni}}, \bibinfo {author} {\bibfnamefont {P.}~\bibnamefont {van~der Smagt}}, \ and\ \bibinfo {author} {\bibfnamefont {M.}~\bibnamefont {Leib}},\ }\href {\doibase 10.1007/s42484-020-00036-4} {\bibfield  {journal} {\bibinfo  {journal} {Quantum Machine Intelligence}\ }\textbf {\bibinfo {volume} {3}},\ \bibinfo {pages} {5} (\bibinfo {year} {2021})}\BibitemShut {NoStop}%
\bibitem [{\citenamefont {McClean}\ \emph {et~al.}(2018)\citenamefont {McClean}, \citenamefont {Boixo}, \citenamefont {Smelyanskiy}, \citenamefont {Babbush},\ and\ \citenamefont {Neven}}]{barrenplateaus}%
  \BibitemOpen
  \bibfield  {author} {\bibinfo {author} {\bibfnamefont {J.~R.}\ \bibnamefont {McClean}}, \bibinfo {author} {\bibfnamefont {S.}~\bibnamefont {Boixo}}, \bibinfo {author} {\bibfnamefont {V.~N.}\ \bibnamefont {Smelyanskiy}}, \bibinfo {author} {\bibfnamefont {R.}~\bibnamefont {Babbush}}, \ and\ \bibinfo {author} {\bibfnamefont {H.}~\bibnamefont {Neven}},\ }\href@noop {} {\bibfield  {journal} {\bibinfo  {journal} {Nature communications}\ }\textbf {\bibinfo {volume} {9}},\ \bibinfo {pages} {4812} (\bibinfo {year} {2018})}\BibitemShut {NoStop}%
\bibitem [{\citenamefont {Wang}\ \emph {et~al.}(2021)\citenamefont {Wang}, \citenamefont {Fontana}, \citenamefont {Cerezo}, \citenamefont {Sharma}, \citenamefont {Sone}, \citenamefont {Cincio},\ and\ \citenamefont {Coles}}]{noiseinducedBI}%
  \BibitemOpen
  \bibfield  {author} {\bibinfo {author} {\bibfnamefont {S.}~\bibnamefont {Wang}}, \bibinfo {author} {\bibfnamefont {E.}~\bibnamefont {Fontana}}, \bibinfo {author} {\bibfnamefont {M.}~\bibnamefont {Cerezo}}, \bibinfo {author} {\bibfnamefont {K.}~\bibnamefont {Sharma}}, \bibinfo {author} {\bibfnamefont {A.}~\bibnamefont {Sone}}, \bibinfo {author} {\bibfnamefont {L.}~\bibnamefont {Cincio}}, \ and\ \bibinfo {author} {\bibfnamefont {P.~J.}\ \bibnamefont {Coles}},\ }\href@noop {} {\bibfield  {journal} {\bibinfo  {journal} {Nature communications}\ }\textbf {\bibinfo {volume} {12}},\ \bibinfo {pages} {6961} (\bibinfo {year} {2021})}\BibitemShut {NoStop}%
\bibitem [{iri(2024)}]{iris_qiskit}%
  \BibitemOpen
  \href {https://qiskit-community.github.io/qiskit-machine-learning/tutorials/02a_training_a_quantum_model_on_a_real_dataset.html} {\enquote {\bibinfo {title} {Training a quantum model on a real dataset},}\ } (\bibinfo {year} {2024})\BibitemShut {NoStop}%
\end{thebibliography}%
\clearpage
\onecolumngrid
\section{Appendix: Optimizing transpilation}
\label{appendix}
\subsubsection{ZZFeatureMap Data Encoding}
\label{sec:ZZFeatureMap}
\begin{figure}[H]
\centering\includegraphics[scale=0.6]{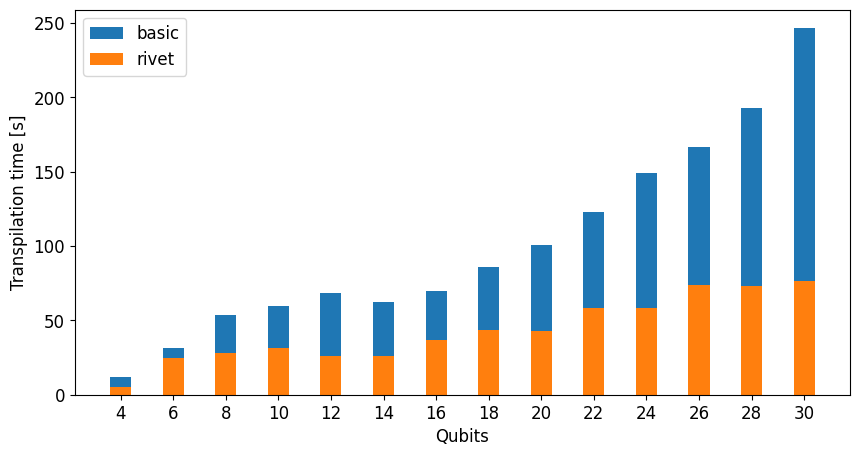}
\caption{Transpilation time for training a PQC using the ZZFeatureMap data encoding with 20 layers and Layerwise Learning approach with 10 layer addition steps of adding 2 layers.}
\label{fig:transpilation_time_ZZfm}
\end{figure}

The ZZFeatureMap is a parameterized quantum circuit used for data encoding, where features are encoded as circuit parameters, and the structure of the circuit captures dependencies between these features through entangling gates. It requires $n$ qubits to encode $n$ features and can capture higher-order correlations by repeating the same circuit multiple times. Figure \ref{fig:transpilation_time_ZZfm} shows the transpilation time for training a PQC with the ZZFeatureMap as the input layer using 20 layers and the Layerwise Learning (LL) approach, with 10 layer addition steps of 2 layers each. The ZZFeatureMap is more transpilation-intensive due to its use of entangling gates, making Rivet a suitable choice. Since the state preparation circuit is parameterized, it can be pre-transpiled once, allowing input features to be appended without recompiling the entire circuit. For different input features, a single pre-transpiled circuit can be reused, unlike amplitude encoding, which requires separate transpilation for each set of features. This results in significant time savings, with Rivet achieving an 8x reduction in transpilation time for 30-qubit circuits.
\end{document}